\documentstyle[twocolumn,aps,prl,epsf,floats]{revtex}
\def\lsi{\raise0.3ex
\hbox{$<$\kern-0.75em\raise-1.1ex\hbox{$\sim$}}}
\def\gsi{\raise0.3ex
\hbox{$>$\kern-0.75em\raise-1.1ex\hbox{$\sim$}}}

\begin{document}
\twocolumn[\hsize\textwidth\columnwidth\hsize\csname
@twocolumnfalse\endcsname

\title{Primordial Hypermagnetic Knots}
\author{ Massimo Giovannini}
\address{{\it Department of Physics and Astronomy, 
Tufts University, Medford, Massachusetts 02155, USA}}

\maketitle

\begin{abstract}
\noindent

Topologically non-trivial 
configurations of the hypermagnetic flux lines lead to the formation 
of hypermagnetic knots (HK) whose decay 
might seed the Baryon Asymmetry of the Universe (BAU).
HK can be dynamically generated  provided
a topologically trivial (i.e. stochastic)
 distribution of flux lines is already present 
in the symmetric phase of the electroweak (EW) theory. In spite of
the mechanism generating the HK, their typical size
must exceed the diffusivity length scale. 
In the  minimal standard model (MSM) (but not necessarily 
 in its supersymmetric extension) HK are washed out. 
A classical hypermagnetic background 
in the symmetric phase of the EW theory can 
produce interesting amounts of gravitational radiation.

\end{abstract}

\vskip1.5pc]

Topologically non-trivial  configurations 
of the magnetic field lines are allowed in terrestrial tokamaks 
and astrophysical
plasmas \cite{mhd}. 
The presence of HK in the 
symmetric phase of the EW theory 
(i.e. $T> T_{c}$)  cannot be theoretically
 excluded. Since the conductivity $\sigma_{c}$ of the EW 
plasma is typically large, in analogy with the electromagnetic case,
 we can expect that the topological structure 
of the hypermagnetic flux lines will be approximately
 conserved (up to corrections of order $1/\sigma_{c}$)  
for sufficiently large scales. 
The importance of the topological properties 
of long range (Abelian) hypercharge magnetic fields has been  
stressed in the  past \cite{vi,ru}.
In \cite{m1} it was argued that if the spectrum of hypermagnetic 
fields is dominated by parity non-invariant Chern-Simons 
(CS) condensates, the BAU could be the result of their decay. 
Most of the mechanisms
often invoked for the origin of large scale magnetic fields in the early 
Universe seem to imply the production of topologically trivial (i.e. 
stochastic) configurations of magnetic fields \cite{seeds}.

The purpose of this Letter is to connect  the topological properties 
of the HK to the generation of the BAU. We show that HK
can be dynamically generated and  can seed the 
BAU only if the correlation scale of the knot is larger than the diffusivity 
scale. We exclude this possibility in the MSM. 
Since hypermagnetic fields present in the symmetric phase of the EW 
theory can radiate gravitational waves (GW),
we propose possible phenomenological tests of our generation 
mechanism.

Suppose that the EW plasma is filled, for $T> T_{c}$ 
with topologically trivial hypermagnetic fields $\vec{\cal H}_{Y}$, 
which  can be physically pictured as a 
collection of flux tubes (closed because of the transversality 
of the field  lines)  evolving independently without 
breaking  or intersecting with each other. If the field
 distribution is topologically 
trivial (i.e. $\langle\vec{\cal H}_{Y} \cdot\vec{\nabla} 
\times\vec{\cal H}_{Y}\rangle =0$) parity is  a  good symmetry 
of the plasma and the field can be completely homogeneous. 
We name hypermagnetic knots those CS condensates carrying 
a non vanishing (averaged)  hypermagnetic helicity
(i.e.  $\langle\vec{\cal H}_{Y} \cdot\vec{\nabla} 
\times\vec{\cal H}_{Y}\rangle \neq 0$). 
If $\langle\vec{\cal H}_{Y} \cdot\vec{\nabla} 
\times\vec{\cal H}_{Y}\rangle \neq 0$  parity is  broken for scales 
comparable with the size of the HK,
 the flux lines are knotted and the field $\vec{{\cal H}}_{Y}$ 
cannot be completely homogeneous.  

In order to seed the BAU a network of HK should be present at high
temperatures \cite{m1,us1}. In fact
for temperatures larger than $T_{c}$
 the fermionic number is stored both in HK 
and in real fermions.  For $T<T_{c}$, 
the HK should release real fermions 
since the ordinary magnetic fields (present {\em after} EW 
symmetry breaking) do not carry fermionic number.
If the EWPT is strongly first order the decay of the HK 
can offer some seeds for the BAU generation \cite{m1}.
This last condition can be met in the 
minimal supersymmetric standard model (MSSM) \cite{PT,mssm}.

Under these hypotheses the integration of the $U(1)_{Y}$ 
anomaly equation \cite{m1}
gives the CS number density carried by the HK
which is in turn related to the density of baryonic number $n_{B}$
for the case of $n_{f}$ fermionic generations \cite{us1}
\begin{equation}
 \frac{n_{B}}{s}(t_{c})=
\frac{\alpha'}{2\pi\sigma_c}\frac{n_f}{s}
\frac{\langle{\vec{{\cal H}}}_{Y}\cdot \vec{\nabla}\times
{\vec{{\cal H}}}_{Y}\rangle}{\Gamma + \Gamma_{{\cal H}}}
\frac{M_{0}\Gamma}{T^2_c},~~\alpha' = \frac{g'^2}{4\pi}
\label{BAU}
\end{equation}
($g'$ is the $U(1)_{Y}$ coupling and $s = (2/45) \pi^2 N_{eff}T^3$ 
is the entropy density; $N_{eff}$ is the
 effective number of massless degrees of freedom  at $T_{c}$
 [$106.75$ in the MSM];
$M_{0}= M_{P}/1.66 \sqrt{N_eff} \simeq 7.1 \times 10^{17} {\rm GeV}$).
In Eq. (\ref{BAU}) $\Gamma$ is the perturbative rate of the 
right electron chirality 
flip processes  (i.e. 
scattering of right electrons with the Higgs and gauge bosons and with 
the top quarks because of their large Yukawa coupling) which 
are the slowest reactions in the plasma and 
\begin{equation}
\Gamma_{{\cal H}} = \frac{783}{22} \frac{\alpha'^2}{\sigma_{c} \pi^2} 
\frac{|\vec{{\cal H}}_{Y}|^2}{T_{c}^2}
\end{equation}
is the rate of right electron dilution induced by the presence of a
 hypermagnetic field. In the MSM we have that 
$\Gamma < \Gamma_{{\cal H}}$ \cite{fl} 
whereas in the MSSM $\Gamma$ can naturally 
be larger than $\Gamma_{{\cal H}}$ \cite{us1}. 
Unfortunately, in the MSM 
a hypermagnetic field can modify the phase diagram of the phase transition 
but cannot make the phase transition strongly first order for large mases of
the Higgs boson \cite {PT2}. 
Therefore, we will concentrate on the case $\Gamma > \Gamma_{\cal H}$ and we
 will show that in the opposite limit the BAU will be anyway small 
even if some (presently unknown) mechanism would make the EWPT strongly 
first order in the MSM.

HK can be dynamically generated.
Gauge-invariance 
and transversality of the magnetic fields suggest
 that perhaps
 the only way of producing  $\langle\vec{\cal H}_{Y} \cdot\vec{\nabla} 
\times\vec{\cal H}_{Y}\rangle \neq 0$ is to 
postulate, a time-dependent interaction between the two (physical) 
 polarizations of the hypercharge field $Y_{\alpha}$.
Having defined the Abelian field strength  $Y_{\alpha\beta} = 
\nabla_{[\alpha} Y_{\beta]}$ and its dual $\tilde{Y}_{\alpha\beta}$ such an 
 interaction can be described, in curved space, by the 
Lagrangian \cite{gar}
\begin{equation}
L_{eff}= \sqrt{-g} \biggl[ 
-\frac{1}{4}Y_{\alpha\beta} Y^{\alpha\beta} + 
c\frac{\psi}{4 M}
Y_{\alpha\beta}\tilde{Y}^{\alpha\beta}\biggr].
\label{action}
\end{equation}
where $g_{\mu\nu}$ is the metric tensor and $g$ its determinant, 
$c$ is the coupling  constant and $M$ is a typical scale.
This  interaction 
is plausible  if the $U(1)_{Y}$ anomaly is coupled, 
(in the symmetric phase of the EW theory ) to  dynamical 
pseudoscalar particles $\psi$ (like the axial Higgs of the MSSM).
Thanks to the presence of pseudoscalar particles, 
the two polarizations of $\vec{{\cal H}}_{Y}$ 
evolve in a slightly different way
producing, ultimately, inhomogeneous HK. 

Suppose that 
an inflationary phase with $a(\tau) \sim \tau^{-1}$  is continuously 
matched, at the transition time $\tau_1$, to a radiation dominated 
phase where $a(\tau)\sim \tau$. Consider then a massive pseudoscalar 
field $\psi$ which oscillates during the last  stages 
of the inflationary evolution
with typical amplitude $\psi_0 \sim M$.
As a result of the inflationary evolution
 $|\vec{\nabla}\psi| \ll \psi'$. Consequently, the 
phase of $\psi$  can  get frozen \cite{linde}. 
Provided  the 
pseudoscalar mass $m$ is larger than the inflationary curvature  scale 
$H_{i}\sim {\rm const.}$, the $\psi$ oscillations are converted, 
at the end of the
 quasi-de Sitter stage, in a net helicity arising as a result 
of the different evolution of the two (circularly polarized) vector potentials
\begin{eqnarray}
&& {Y}_{\pm}'' + \sigma Y'_{\pm}+
\omega_{\pm}^2 {Y}_{\pm} =0,~~\vec{H}_{Y} = \vec{\nabla} 
\times \vec{Y}
\nonumber\\
&&\psi\sim a^{-3/2} \psi_0\sin{[m (t- t_1)]},~~~
\omega^2_{\pm} = k^2 \mp k \frac{c}{M} a \dot{\psi}    
\label{Y}
\end{eqnarray}
(where we denoted with $\vec{H}_{Y} = a^2 {\vec{\cal H}}_{Y}$  the 
curved space fields and with $\sigma= \sigma_c a$ the rescaled 
hyperconductivity; the prime denotes derivation with respect to 
conformal time $\tau$ whereas the over-dot denotes differentiation 
with respect to cosmic time $t$). 
Since $\omega_{+}\neq \omega_{-}$ the helicity 
gets amplified according to Eq. (\ref{Y}) 
and the BAU can be obtained (for $\Gamma > \Gamma_{{\cal H}}$) 
from Eq. (\ref{BAU}) 
\begin{eqnarray}
&& \frac{n_{B}}{s} = \delta \biggl(\frac{m}{H_{i}}\biggr)^5 
\biggl( \frac{H_{i}}{M_{P}}
\biggr)^{\frac{5}{2}} e^{c (\frac{m}{H_{i}}) (\frac{\psi_{0}}{M})}
e^{-2(\frac{\omega_{{\rm m}}}{\omega_{\sigma}})^2}
\nonumber\\
&&\delta = \frac{45 g'^2 c^5 n_{f} 
N^{\frac{1}{4}}_{eff}}{512 \pi^6 \sigma_0}\frac{M_0}{T_{c}},~~
\sigma_0 = \frac{\sigma_{c}}{T_c},
\label{BAU2}
\end{eqnarray}
where $\omega_{{\rm m}}= k_{{\rm m}}/a\sim (c/2)(\psi_0/M) m $ 
is the maximally 
amplified frequency corresponding to the center of the 
 first (and larger) instability band of the Mathieu-type
 equation for $Y_{\pm}$  and 
$\omega_{\sigma}(\tau_c) \sim \sigma_0^{1/2} N_{eff}^{1/4} (T_c/M_{P})^{1/2} 
T_c$
is the maximal (hyperconductivity) frequency of the spectrum. 
The possible  oscillations arising in 
$\langle\vec{\cal H}_{Y} \cdot\vec{\nabla} 
\times\vec{\cal H}_{Y}\rangle$ are smeared out
as a consequence of the growth of the 
hyperconductivity which is exactly zero 
in the inflationary phase but which gets large as soon as the 
radiation phase is approached. 

\begin{figure}
\vspace*{0cm}
\epsfxsize = 6 cm
\centerline{\epsffile{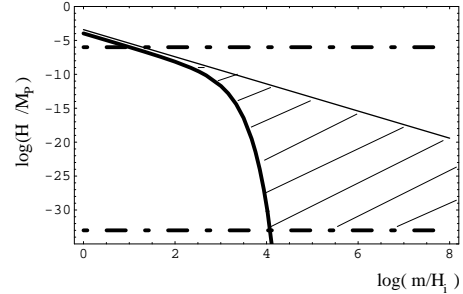}}
\caption[a]{With the full thick line we illustrate the bound of 
Eq. (\ref{c1}) for a fiducial set of parameters
($c=0.01$, $\sigma_0\sim 70$ and $N_{eff} = 106.75$). 
In order to 
produce a sizable BAU we have to be within the shaded area. The thin  
line corresponds to the hyperconductivity 
bound (i.e. $k<k_{\sigma}$) for Fourier modes amplified during 
the inflationary epoch and evolving, subsequently, in the radiation phase.} 
\label{fl1}
\end{figure}

Without fine-tuning the amplitude of the $\psi$ oscillations 
we are led to require $\psi_0\sim M$.
If we impose that $n_{B}/s \gsi 10^{-10}$ we get, from Eq. (\ref{BAU2}) and in 
the case of three fermionic generations, the condition
\begin{eqnarray} 
\log_{10}{\frac{H_i}{M_{P}}}\gsi 
-8.5 +\log_{10}{
\biggl[\frac{\sigma_0^{\frac{2}{5}}}{c^2~N^{\frac{1}{10}}_{eff}}\biggr]} 
\nonumber\\
- \frac{2c}{5} \frac{m}{H_{i}} \log_{10}{e} 
- 2 \log_{10}{\frac{m}{H_{i}}},
\label{c1}
\end{eqnarray}
illustrated in Fig. \ref{fl1} with the thick (full) line.
In order to produce a sizable 
BAU ($ \gsi 10^{-10}$ ) we need to be above the thick full line
 but also below the thin (full) line representing the condition 
$\omega_{{\rm m}}(t)<\omega_{\sigma}(t)$. Moreover, in order to be
consistent with the (undetected) tensor contribution to the Cosmic 
Microwave Background anisotropy we are led to require 
$H_{i}/M_{P} \lsi 10^{-6}$. In order to have inflation prior to 
the onset of the EW epoch we must impose $H_{i}/M_{P} > 10^{-33}$.
In Fig. \ref{fl1} this last requirement corresponds
to the shaded region  within the two dot-dashed lines. Thus, provided 
$m/H_{i} \gsi 10^{4}$ the pseudoscalar oscillations produce sufficient 
helicity to seed the BAU also for reasonably small inflationary scale
$H_{i} \sim 10^{-22} M_{P}$ (see \cite{us2} for further details).

During an inflationary stage $\sigma\rightarrow 0$.
If the $\psi$ oscillations take place in a radiation dominated epoch
($\sigma \neq 0$) 
the evolution of the hypercharge are damped, from the very beginning, 
thanks to the finite value of the hyperconductivity 
according to
\begin{equation}
\sigma Y_{\pm}' + \biggl[  k^2 \mp k c \frac{\psi'}{M}\biggr]Y_{\pm} =0.
\label{coneq1}
\end{equation}
More precisely, for $T>T_{c}$  Eq. (\ref{coneq1}) should be 
complemented by the equations of  anomalous magnetohydrodynamics (AMHD)
 accounting for  the coupled evolution 
of $\psi$, $\mu_{R}$ (the right electrons chemical potential) 
and of the velocity field $\vec{v}$
\begin{eqnarray}
&&\frac{(\mu_R a)'}{a}  = - \frac{g'^2
}{4\pi^2} \frac{783 }{88} 
\frac{{\vec{H}}_{Y}\cdot{\nabla\times\vec{H}}_{Y}  }{\sigma a^3 T^3} 
\nonumber\\
&&- (\Gamma +\Gamma_{H}) (\mu_{R} a) + D_{R} \nabla^2 (\mu_{R}a),
\label{mu}\\
&&{{\vec{H}}_{Y}}' =- \frac{4 a
\alpha'}{\pi\sigma}  
\vec{\nabla}\times\left({\mu_{R}\vec{H}}_{Y}\right) 
\nonumber\\
&&-\frac{c}{M} \vec{\nabla}\times [ \psi' {\vec{H}}_{Y} ]
+{\vec{\nabla}}\times(\vec{v}\times{\vec{H}}_{Y})
+ \frac{1}{\sigma}
 \nabla^2 {\vec{H}}_{Y},
\label{hyperdiff}\\
&&\vec{v}' + [\vec{v}\cdot\vec{\nabla}]\vec{v} =
\frac{[\vec{H}_{Y}\cdot\vec{\nabla}]\vec{H}_{Y}}{[\rho + p]} +
 \nu \nabla^2\vec{v},
\label{navier}
\end{eqnarray}
where $a(\tau) d\tau = dt$ and $H =(\ln{a})^{\cdot}$.
In Eq. (\ref{navier}) we neglected the 
Lorentz term which is subleading in the case of maximally 
helical fields \cite{gar} and we also used the incompressible 
closure (i.e. $\vec{\nabla}\cdot\vec{v}=0$) of the AMHD equations in the 
assumption of a perfect fluid with radiation-like equation of state
$p=\rho/3$. In Eq. (\ref{mu}) on top of the chirality 
changing rates we introduced the diffusion coefficient of the right 
electron chemical potential $D_{R}$ leading to 
a typical diffusion scale $k_{D} \sim \alpha' (T/M_0)^{1/2} T$. 
Eqs. (\ref{mu})--(\ref{navier}) should be generalized to include, in principle,
{\em all} the processes which are in local thermal equilibrium for $T>T_{c}$.
However, if we want to focus our attention 
 on the generation of HK right before the EWPT, 
we are led to consider with special care the right-electrons
whose equilibration temperature can fall (in the MSM) in the TeV range
\cite{fl}.
If the thermal and hypermagnetic diffusion 
coefficients are of the same order (i.e. $\nu\sim \sigma$),
the solution of Eq. (\ref{coneq1}) together with 
Eqs. (\ref{mu})--(\ref{hyperdiff}) determines the evolution of 
the HK at finite fermionic density.
If we insert the result into Eq. (\ref{BAU}) 
we  get, in the limit $k\ll k_{\sigma}$ and $k\ll k_{D}$, 
that  the BAU is given by 
\begin{equation}
\frac{n_{B}}{s} \simeq \frac{45 n_f }{8\pi^3 \sigma_0} ~c~ \alpha'
 \frac{\Delta\psi}{M} 
r,~~~\omega_{{\rm m}} = \frac{c}{2 a} 
\frac{\Delta\psi}{M} \frac{T_c^2}{M_{0}},
\label{BAUrad}
\end{equation}
where $r = |{\vec{\cal H}_{Y}}|^2/ ( N_{eff} T_{c}^4)$ 
is the critical fraction of energy density stored 
in the initial (topologically trivial) hypermagnetic distribution
for $\omega\sim \omega_{\rm m}$.
Notice also that $\omega_{{\rm m}}$ differs,
 in the present case, from the maximally amplified frequency defined 
in the context of the inflationary amplification. 
If we do not fine-tune the initial amplitude of the 
oscillations to be much larger than $M$ we have that $\Delta\psi \sim M$.
Concerning Eq. (\ref{BAUrad}) three remarks are in order:

i) it holds provided $\omega_{{\rm m}} <\omega_{\sigma}$ (indeed
 only in this limit Eq. (\ref{BAU}) is meaningful \cite{us1});

ii) it holds  provided 
we are in the context MSSM since only in this 
case $\Gamma$ can be large enough and  EWPT can be 
strongly first order;

iii) it  can give a relevant BAU if (and only if)
 an initial distribution of topologically trivial
hypermagnetic fluctuations is postulated, namely 
 $r(\omega_{\rm m})$ needs 
to be at least $10^{-3}$, whereas, in the case of vacuum fluctuations 
$r(\omega_{{\rm m}}) \sim N_{eff} (\omega_{{\rm m}}/T)^4 \sim 10^{-33}$.
This last point implies, physically, that the BAU cannot be generated from
vacuum fluctuations of the hypercharge fields.
 Our observations are not in agreement  with \cite{br} where it is assessed
 that the correct value of the BAU can be reproduced in the MSM, 
 for $k_{{\rm m}}\sim~ T$ and disregarding the role of the slowest 
processes in the EW plasma. According to our present analysis we do not share 
these last statements.

It  seems instead more natural, in our scenario, 
to assume that at the scale where $\psi$
oscillates in the radiation epoch 
there is a topologically trivial (stochastic) hypermagnetic
distribution  
since it might have been  generated, for example, thanks to the breaking of 
conformal invariance or through some other mechanism \cite{seeds}.
We focus our attention on temperatures in the TeV range where the 
right electrons can be still out of thermal equilibrium. Inspired
by the axial Higgs we will be concerned with pseudoscalar masses 
$m\gsi 300$ GeV as required in order to have a MSM Higgs
 sector not too different from the one of the MSSM \cite{hun}. 

\begin{figure}
\vspace*{0cm}
\epsfxsize = 6 cm
\centerline{\epsffile{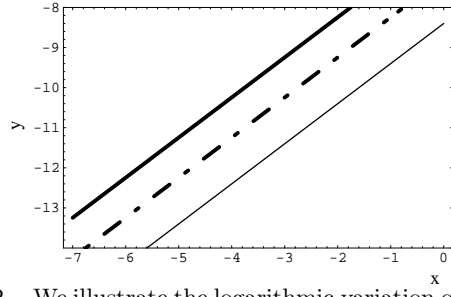}}
\caption[a]{ We illustrate the logarithmic variation of the BAU computed 
in Eq. (\ref{BAUrad2}) for different sets of parameters. More precisely
we have $c=0.01$, $\sigma_0 =70$ (full thick line), $c=10^{-3}$, 
$\sigma_0=70$ (dot-dashed line), $c=10^{-4}$, $\sigma_0=100$ 
(full thin line).}
\label{fl2}
\end{figure}
Comparing the BAU obtained from the previous equation with $10^{-10}$ we 
obtain a condition on the parameters of the model, namely,  by 
imposing $y \gsi -10$
\begin{equation}
- 2.4 + \log_{10}{c} - \log_{10}{\sigma_0} 
+ \log_{10}(\frac{\Delta\psi}{M}) + x \gsi -10 
\label{BAUrad2}
\end{equation}
[where now $x = \log_{10}{r(\omega_{{\rm m}})}$ and $y= 
\log_{10}{(n_{B}/s)}$].
This result is illustrated in Fig. \ref{fl2} for the accessible 
region of the parameter space in the case $n_{f}=3$.
One could also wonder  if the MSM physics would be enough
to produce HK. Let us assume, for a moment, the MSM. 
Then, as argued in \cite{us1,us2}, 
$\Gamma_{{\cal H}}>\Gamma$. Even assuming
the EWPT to be strongly first
order in the presence of a large hypermagnetic field (which is 
not the case) from Eq. (\ref{BAU}) the BAU will be given 
by the expectation value of the highly non-local operator
\begin{equation}
\langle \frac{\vec{{\cal H}}_{Y}\cdot\vec{\nabla}
\times\vec{{\cal H}}_{Y}}{|\vec{{\cal H}}_{Y}|^2}
\rangle \simeq 
\frac{\langle\vec{{\cal H}}_{Y}\cdot\vec{\nabla}\times\vec{{\cal H}}_{Y}
\rangle}
{\langle|\vec{{\cal H}}_{Y}|^2\rangle}, 
\end{equation}
where the last equality can be obtained for sufficiently large scales 
\cite{us1}. Then, using Eq. (\ref{coneq1}) the BAU turns out to be 
\begin{equation}
\frac{n_{B}}{s} \simeq 0.04~c~ \biggl(\frac{\Delta\psi}{M}\biggr)
 \biggl(\frac{T_{R}}{M_0}\biggr).
\label{BAU3}
\end{equation}
where $T_{R}\sim 80$ TeV is the right-electron equilibration temperature 
\cite{fl}. For the accessible region of the parameter space 
 the BAU is of the order of $10^{-18}$.

Our considerations can have direct phenomenological implications. 
It is amusing to notice that if
a hypermagnetic background is present for $T> T_c$, then, as discussed
in \cite{mmm} in the context of ordinary MHD,  the energy momentum tensor 
will acquire a small anisotropic component which will source the evolution 
equation of the tensor fluctuations $h_{\mu\nu}$ of the metric $g_{\mu\nu}$: 
\begin{equation}
h_{ij}'' + 2 {\cal H} h_{ij}' - \nabla^2 h_{ij} = - 16 \pi G
\tau^{(T)}_{ij}.
\label{GWeq}
\end{equation}
where $\tau^{(T)}_{ij}$ is the {\em tensor} component of the 
{\em energy-momentum tensor} \cite{mmm} 
of the hypermagnetic fields. Suppose now, as assumed in \cite{PT2} that 
$|\vec{{\cal H}}|$ has constant amplitude and that it is also 
homogeneous. Then 
as argued in  \cite{rg} we can easily deduce 
the critical fraction of energy density  present today in relic gravitons 
of EW origin 
\begin{equation}
\Omega_{\rm gw}(t_0) = \frac{\rho_{\rm gw}}{\rho_c} 
\simeq z^{-1}_{{\rm eq}}
r^2,~~\rho_{c}(T_{c})\simeq N_{\rm eff} T^4_{c }
\end{equation}
($z_{\rm eq}=6000$ is the redshift from the time of matter-radiation,
 equality to the present time 
$t_0$). Because of the structure of the AMHD equations, stable 
hypermagnetic fields will be present not only for 
$\omega_{\rm ew}\sim k_{\rm ew}/a$ but 
for all the range $\omega_{{\rm ew}} <\omega< \omega_{\sigma}$. Let us assume, 
 for instance, that $T_{c} \sim 100 $ GeV and $N_{eff} = 106.75$. 
Then, the (present) values of 
$\omega_{\rm ew}$ and $\omega_{\sigma}$ will be, respectively, 
$2\times 10^{ -5} $ Hz and $ 1.5 \times 10^{3}$ Hz. Suppose now, 
as assumed in \cite{PT2} that $|{\vec{\cal H}}|/T_{c}^2\gsi 0.3$. Thus 
$r\simeq 0.1$--$0.01$ and $h_0^2 \Omega_{\rm GW} \simeq 
10^{-7}$--$10^{-8}$ which is a respectable signal compatible with (but smaller 
than) the bounds coming from BBN \cite{mgas} and implying 
$h_0^2 \Omega_{\rm GW} \lsi 10^{-6}$ \cite{note}.

The author is deeply indebted to M. Shaposhnikov for very valuable discussions 
and collaboration.


\begin{thebibliography}{99}

\bibitem{mhd} N. A. Krall and A. W. Trivelpiece, {\it Principles of
Plasma Physics}, (San Francisco Press, San Francisco 1986);  
E. N. Parker, Astrophys. J. {\bf 122}, 293 (1955);
{\em ibid}. {\bf 163}, 252 (1971).

\bibitem{vi} A. Vilenkin, Phys. Rev. D {\bf 22}, 3067 (1980).

\bibitem{ru} V. Rubakov and A. Tavkhelidze, Phys. Lett. {\bf B 165},
109 (1985); V. Rubakov, Prog. Theor. Phys. {\bf 75}, 366 (1986); 
A. N. Redlich and L. C. R. Wijewardhana, Phys. Rev.
 Lett. {\bf 54}, 970 (1984).

\bibitem{m1}  M. E. Shaposhnikov, JETP Lett. {\bf 44}, 465 (1986);
Nucl. Phys. B {\bf 287}, 757 (1987);{\it ibid.} {\bf  299}, 797
(1988).

\bibitem{seeds} T. Vachaspati, Phys. Lett. {\bf B 265}, 258 (1991);
 K. Enqvist, P. Olesen, Phys. Lett. {\bf B 319}, 178 (1993);
T. W. Kibble and A. Vilenkin, Phys. Rev. {\bf D 52}, 679 (1995);
 G. Baym, D. Bodeker and L. McLerran Phys.Rev. {\bf D 53}, 662 (1996);
M. Joyce and M. Shaposhnikov, Phys. Rev. Lett. {\bf
79}, 1193 (1997);M.S. Turner and L.M. Widrow, Phys. Rev.
{\bf D 37}, 2743 (1988); B. Ratra,  Astrophys. J. Lett, {\bf 391}, L1
(1992); A. Dolgov and J. Silk, Phys. Rev. {\bf D 47}, 3144 (1993);
M. Gasperini, M. Giovannini and G. Veneziano, Phys.
Rev. Lett. {\bf 75}, 3796 (1995); Phys. Rev. {\bf D 52}, 6651 (1995).

\bibitem{us1}  M. Giovannini and M.  Shaposhnikov, Phys. Rev. D {\bf 57}, 
2186 (1998); M. Giovannini and M. Shaposhnikov, Phys. Rev. Lett. {\bf 80},
 22 (1998).

\bibitem{PT} K. Kajantie, M. Laine, K. Rummukainen and
M. Shaposhnikov, Nucl. Phys. {\bf B 495}, 413 (1997).

\bibitem{mssm} M. Carena, M. Quiros and C. Wagner, Phys. Lett. B {\bf 380}, 
81 (1996); M. Laine Nucl. Phys. B {\bf 481}, 43 (1996); 
J. Cline and K. Kainulainen, Nucl. Phys. B {\bf 482}, 73 (1996).

\bibitem{fl} B. Campbell, S. Davidson, J. Ellis and K. Olive, Phys.
Lett. {\bf 297B }, 118, (1992); L.~E. Ibanez and F.~Quevedo,
Phys. Lett., {\bf B283}, 261, 1992; J. M. Cline, K. Kainulainen and K.
 A. Olive, Phys. Rev. Lett. {\bf 71}, 2372 (1993); Phys. Rev. {\bf D 49}, 6394
(1993).

\bibitem{PT2} K. Kajantie, M. Laine, J. Peisa, K. Rummukainen,
 and M. Shaposhnikov, Nucl.Phys.B {\bf 544}, 357 (1999).

\bibitem{gar} W. Garretson, G. Field, and S. Carroll, Phys. Rev. D {\bf 46}, 
5346 (1992); S. Carroll and G. Field, {\em ibid}. {\bf 43}, 3789 (1991); 
S. Carroll and G. Field, astro-ph/9811206. 

\bibitem{linde} A. D. Linde and D. H. Lyth, Phys. Lett. B {\bf 246},
 353 (1990); A. D. Linde, Phys. Lett. B {\bf 259}, 38 (1991).

\bibitem{us2} M. Giovannini, in preparation.

\bibitem{br} R. Brustein and D. Oaknin, Phys. Rev. Lett. 
{\bf 82}, 2628 (1999). 

\bibitem{hun} K. Sasaki, M. Carena, and C. Wagner, Nucl. Phys. B {\bf 381},66
 (1992); M. Carena, J. Espinosa, M. Quir\'os, and C. Wagner, Nucl. Phys. 
B {\bf 461}, 407 (1996).

\bibitem{mmm} M. Giovannini, Phys. Rev.D {\bf 58}, 124027 (1998). 

\bibitem{rg} D. Deryagin, D. Grigoriev, V. Rubakov and M. Sazhin, 
Mod. Phys. Lett. A {\bf 11}, 593 (1986). 

\bibitem{mgas} M. Gasperini and M. Giovannini, Phys. Rev. D {\bf 47}, 
1519 (1993).

\bibitem{note} Notice that the the pulsar timing bound (which applies 
for present frequencies $\omega_{P}\sim 10^{-8}$ Hz and implies 
$h^2_{0}\Omega_{\rm GW} \lsi 10^{-8}$) is automatically satified since
our hypermagnetic background is defined for $10^{-5} {\rm Hz} 
\lsi \omega \lsi 10^{3}~{\rm Hz}$. 
\end{thebibliography}
\end{document}